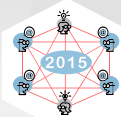

# Invited Abstract: A simulation package for energy consumption of content delivery networks (CDNs)

Mohammadhassan Safavi     Saeed Bastani
Department of Electrical and Information Technology, Lund University, Lund, Sweden
{*mohammadhassan.safavi, saeed.bastani*}@eit.lth.se

*Abstract*—Content Delivery Networks (CDNs) are becoming an integral part of the future generation Internet. Traditionally, these networks have been designed with the goals of traffic offload and the improvement of users' quality of experience (QoE), but the energy consumption is also becoming an indispensable design factor for CDNs to be a sustainable solution. To study and improve the CDN architectures using this new design metric, we are planning to develop a generic and flexible simulation package in OMNet++. This package is aimed to render a holistic view about the CDN energy consumption behaviour by incorporating the state-of-the-art energy consumption models proposed for the individual elements of CDNs (e.g. servers, routers, wired and wireless links, wireless devices, etc.) and for the various Internet contents (web pages, files, streaming video, etc.).

## I. INTRODUCTION

With the tremendous growth of the Internet traffic and the demand for guaranteed quality of service, the content delivery networks (CDNs) have emerged as a promising solution to offload the traffic and to reduce content access delay, so that both content providers and consumers benefit directly. There is also a new emerging design goal for CDNs, motivated by the present trend of energy consumption in the ICT sector which implies for an urgent need for re-assessing and re-designing Internet systems to achieve energy efficient architectures. In particular, the content delivery networks can play a significant role in harnessing energy consumption per bit delivered to the end users. This is supported by the fact that the Internet is experiencing an unprecedented growth in the type and volume of contents. The most significant phenomenon of this type is video traffic. A recent trend of the Internet usage indicates an increased demand for streaming media (IPTV, Youtube etc.) in the downlink as well as an increased user generated traffic (upload of video to Facebook, Youtube, Vimeo etc.) in the uplink side. Knowing the fact that the delivery of video contents is highly bandwidth and processor demanding, it is of prominent value to understand how the current CDNs behave in terms of energy efficiency, and how to calibrate the existing CDNs or to design new ones by taking into account the energy efficiency as an additional design metric. The study of energy efficiency of the existing CDNs and, more importantly, the design of new CDNs require that the contributions of a tremendous number of factors to the energy consumption is taken into consideration. On one side, the end-users are increasingly inclined to use battery powered mobile devices or portable computers. On the other side, there are a huge number of network elements (e.g. servers, storages, routers, links, etc.) responsible for providing, managing, and carrying the Internet contents to the users. Therefore, the performance study of a given CDN architecture requires that the entire set of network elements with a role in carrying contents from a content provider to the consumers are taken into account. Moreover, contents have different types and properties and users have different preferences over contents and use different devices to access those contents. With this extremely huge set of elements involved in a CDN, simulation as a tool appears as the most affordable way of studying the energy efficiency of the existing CDNs, and to design new CDNs with respect to energy efficiency metric.

In the following, we present the building blocks of an envisioned simulation package for the studying of energy consumption behaviour of CDNs.

## II. A SIMULATION PACKAGE FOR ENERGY CONSUMPTION BEHAVIOUR OF CDNs

Our proposed simulation package is depicted in Figure 1. The description of the building blocks of the proposed package is as follows:

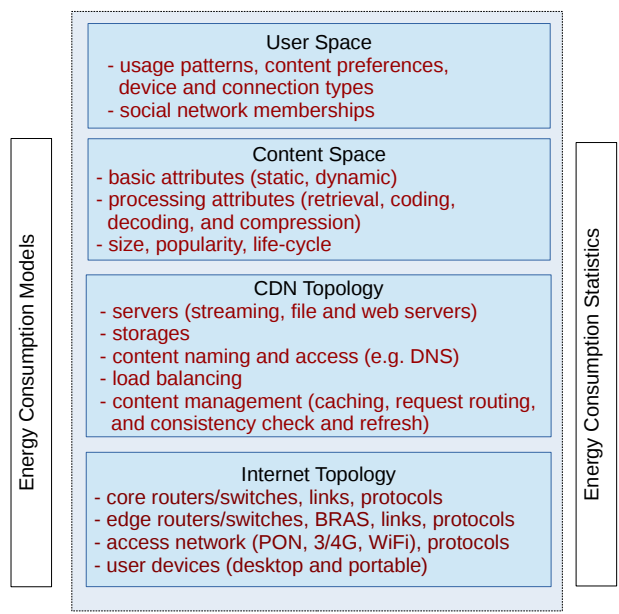

Fig. 1: Building blocks of the proposed simulation package

- **User Space:** this module defines the various attributes of a user including the frequency and time patterns of user access to contents of various types. Also, it defines the type and properties of the device(s) used by the user and the way they connect to the network. The device and connection type determine the amount of energy consumed in the user premise to access a unit of content. Additionally, the user's social activities are used to predict the frequency by which the user will access the contents shared by the members of her social network. This latter information is used by the content management component of the CDN to efficiently cache and replicate those contents deemed to become popular.
- **Content Space:** it defines the contents attributes including the set of content types and their (relative) popularities extracted from field data or described by appropriate distribution functions. It also defines the life-cycle of each content characterized by the time period the content is present in the network and its popularity pattern during its lifetime. Another attribute of a content which directly relates to energy consumption is the resources (bandwidth and processor time) required to carry the content from its source to the user device. Apparently, this property depends on factors other than the content itself, including the types and properties of the network elements involved in processing and carrying the content. However, in this module the focus is on the abstract features of the content, e.g. required bitrate and the coding and decoding complexity of the (video) content regardless of the hosting devices.
- **CDN Topology:** this module defines the network topology and the enabling functions/tasks of the CDN. Two different groups of attributes are specified in this module: the hardware and software. The hardware attributes defines the the type, capacity, and the power density of the servers and the storage devices. Other hardware elements are defined in this group including the networking equipment (switches) and links to interconnect the servers and storages internally and externally with the rest of the network. The software attributes are specified by the content management tasks (as a major task of the CDN), content addressing and request routing, and load balancing function.
- **Internet Topology:** it defines the network equipment and the interconnection devices and protocols, altogether representing the Internet as the underlying network for CDNs.
- **Energy Consumption Statistics:** this module implements statistic collection functions and the GUIs needed to demonstrate and analyse the energy consumption behaviour by the individual network layers, and the energy efficiency of the CDN in its entirety.
- **Energy Consumption Models:** this module is regarded as the core part of the envisioned simulation package. It relies on the energy consumption models proposed in the literature. These models cover a wide range of CDN elements from processing units, cooling equipment, storage, and communication interfaces. This module takes as input those features of a typical content which describes, for example, the time complexity of the content processing (e.g. streaming, coding, decoding), and combines them with the power density of a given network element (e.g. a processing unit of a server) currently hosting the content, and calculates the amount of energy consumed to perform the required operation on the content.

In the following, we briefly describe some specific considerations about the CDN architecture subject to implement in the simulation package and the energy consumption models to be incorporated in this package.

In the traditional CDN architecture, there is a single set of surrogate (or delivery) servers located in a single architectural level, but in the recent CDNs the surrogate servers are located in multiple levels to render a hierarchical architecture [1] (see Figure 2). Due to the generalization concerns, we will concentrate on this reference architecture in our simulation package.

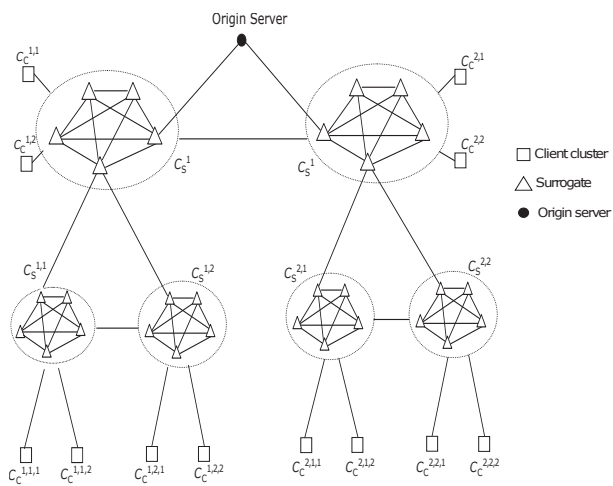

Fig. 2: A state-of-the-art CDN model

In the reference CDN (Figure 2), $C_S$ is server cluster and $C_C$ is client cluster, and the superscripts refer to the index of the individual clusters and systems.

It is ideal to have end-to-end energy consumption models in place and incorporate them in the simulation package. However, such models have not been developed for the reference CDN architecture depicted in Figure 2, and the existing models each describes a subset of the elements in the end-to-end path from the source to the user. Therefore, we opt to select some appropriate models of this kind and integrate them into a holistic end-to-end energy consumption model. The most complete model so far has been proposed in [2], which describes the energy consumption of an IPTV network. In this macroscopic model, the energy consumption in watts-hour for

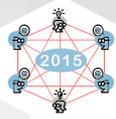

downloading video contents is expressed as follows:

$$E_{download} = 4\frac{B}{3600}\left(\frac{3P_{ES}}{C_{ES}} + \frac{P_G}{C_G} + \frac{2P_{PE}}{C_{PE}} + \frac{(H+1)P_C}{C_C} + \frac{HP_{WDM}}{C_{WDM}} + \frac{P_{SR}}{C_{SR}}\right) + 2\frac{BR}{D}\left(\frac{P_{SD}}{S_{SD}}\right) \quad (1)$$

where $P$ is the power density and $C$ is the capacity of the given device. The subscripts $ES$, $G$, $PE$, $C$, $WDM$, and $SR$ represent Ethernet switches, gateway routers, provider edge routers, core routers, WDM equipment, and content servers, respectively. The parameters $H$ and $B$ respectively indicate the video content size and the average number of hops between the content server and the end-user(s). $R$ and $D$ represent the number of replicated copies of a content in the data centres and the average number of downloads per hour. From the model, it is evident that with the number of hops $H$ increasing, the energy consumption also increases. For this reason, surrogate servers are deployed with a suitable density to cache the contents and to reduce the distance between the content server and the end-user(s) [3]. The model described above suffers from some limitations, including the lack of differentiation between different access technologies and the absence of specialized models to describe the energy consumption of different content types. To overcome the former limitation, we propose to augment the model with dedicated energy consumption models proposed for wireless access as the most prevalent access technology and the most energy consuming part of the network [4]. In view of this, our first attempt is to incorporate an energy consumption model proposed by [5] into the simulation package. This model describes the power consumption of an 802.11 device by a concise expression as follows:

$$P = \rho_{id} + \rho_{tx}\tau_{tx} + \rho_{rx}\tau_{rx} + \gamma_{xg}\lambda_g + \gamma_{xr}\lambda_r \quad (2)$$

where $\rho_{id}$ is the power consumption in idle mode, $\rho_{tx}$ is transmission power coefficient, $\rho_{rx}$ is reception power coefficient, $\tau_{tx}$ is packet transmission airtime, $\tau_{rx}$ is packet reception airtime, $\lambda_g$ is packet generation rate, $\lambda_r$ is packet reception rate, $\gamma_{xg}$ is the energy toll associated to the processing of each individual frame at the transmitting device, and $\gamma_{xr}$ is the energy toll associated to the processing of each individual frame at the receiving device. The model described above is simple and straightforward to implement in OMNet++, e.g. by integrating it into the Energy Framework and the wireless access modules of the INET simulation framework.

The models described by expressions 1 and 2 are mainly focused on the transmission energy, and simplify the processing energy to a great extent. In these models, the packets are regarded as raw data units without taking into account the specific processing features of the given content type. This approach, however, leads to a significant inaccuracy of energy consumption modelling of processor demanding contents such as video streaming. Therefore, our major task in developing the envisioned simulation package is to fill the mentioned gap by using content-specific energy consumption models. Our initial focus in this respect is video contents. Along this line, video coding, decoding and streaming account for the major processing factors of video contents, though these factors are not symmetric in terms of processing complexity and energy consumption behaviour. While video coding is more processor demanding than the decoding, the latter is invoked more frequently (per user request), and thus, from a CDN perspective, it is the dominant factor. An example model of decoding energy consumption is proposed in [6], where the authors derive an energy consumption model from the time complexity of a given codec standard (here H.264 /MPEG 4 AVC). It is generally possible to map from time complexity to energy consumption, on condition that the type of the coding/decoding algorithm (i.e. serial, concurrent, multi-threading) and the speed and architecture of the host processor (single and multi-core) are known. There are also other energy consumption parameters beyond coding/decoding, such as size, resolution and luminance settings of the user's device LCD. The intensity of memory access (i.e. space complexity) is another important factor required to be captured in the desired model.

In conclusion, we believe that OMNet++ yields a high level of flexibility in implementing the proposed simulation package for energy consumption of CDNs. The usefulness of such a package for the networking researchers is substantially dependent on the availability of accurate energy consumption models proposed in the literature. As mentioned previously, there are some promising models such as those described by expressions 1 and 2, which can be incorporated in the first roll-out of the envisioned simulation package. There are also microscopic models, such as [6], which are capable of capturing the essential features of video processing and mapping it to the outcome energy consumption behaviour.

ACKNOWLEDGEMENTS

This work is funded by the European Celtic-Plus project CONVINcE.

ACKNOWLEDGEMENTS

This work is funded by the European Celtic-Plus project CONVINcE.